\documentstyle[preprint,tighten,aps]{revtex} % Specifies the document style.
\begin{document}           %
\draft
\preprint{\vbox{\noindent
          \null\hfill MIT-CTP\#2357\\
          \null\hfill hep-ph/9412246\\
          \null\hfill INFNCA-TH-94-16}}
\title{Constraining Spectral Functions\\
 at Finite Temperature and  Chemical Potential \\
with Exact Sum Rules in Asymptotically Free Theories}

\author{Suzhou Huang$^{(1,2)}$\cite{email}
        and Marcello Lissia$^{(1,3)}$\cite{email} }
\address{
$^{(1)}$Center for Theoretical Physics, Laboratory for Nuclear Science
and Department of Physics, \\
Massachusetts Institute of Technology, Cambridge, Massachusetts 02139\\
$^{(2)}$Department of Physics, FM-15, University of Washington,
Seattle, Washington 98195~\cite{present}\\
$^{(3)}$Istituto Nazionale di Fisica Nucleare,
via Ada Negri 18, I-09127 Cagliari, Italy~\cite{present}\\
and Dipartimento di Fisica dell'Universit\`a di Cagliari,
I-09124 Cagliari, Italy
         }
\date{\today}
\maketitle                 % Produces the title.
\begin{abstract}
Within the framework of the operator product expansion (OPE)
and the renormalization group equation (RGE), we show that
the temperature and chemical potential dependence of the
zeroth moment of a spectral function (SF) is completely
determined by the one-loop structure in an asymptotically
free theory, and in particular in QCD.
Logarithmic corrections are found to play an essential
role in the derivation. This exact result constrains the shape
of SF's, and implies striking effects near phase transitions.
Phenomenological parameterizations of the SF, often used in
applications such as the analysis of lattice QCD data or QCD
sum rule calculations at finite temperature and baryon density
must satisfy these constraints. We also explicitly illustrate
in detail the exact sum rule in the Gross-Neveu model.
\end{abstract}
\maketitle
\widetext
\section{Introduction}
Hadronic properties at finite temperature and baryon density are
of great importance in the phenomenology of heavy ions collisions,
star interior and the early universe. Moreover, the theoretical
expectation of transitions to a chirally symmetric phase and,
perhaps, to a quark-gluon plasma phase contributes to the interest
in studying the effect of matter and temperature on the
quantum chromodynamics (QCD) vacuum.

  Our present understanding of QCD at finite temperature ($T$) and
baryon density (or chemical potential $\mu$) is mainly limited
in the Euclidean realm, due to the lack of non-perturbative and
systematic calculating tools directly in the Minkowski space.
Typical methods, with QCD Lagrangian as the starting point, are
the OPE and lattice simulations. Because these two formulations
are intrinsically Euclidean, only static quantities are conveniently
studied. In order to gain dynamical informations, which are more
accessible experimentally, the analytic structure implemented
through dispersion relations often have to be invoked within the
theory of linear response.

In principle, dispersion relations allow the determination of the
spectral function (SF), which carries all the real-time information,
from the corresponding correlator in the Euclidean region. In practice,
realistic calculations, e.g. OPE or lattice simulations, yield only
partial information on the correlators, making impossible a direct
inversion of the dispersion relation.

Therefore, the standard approach assumes a phenomenological motivated
functional form with several parameters for the SF, and uses
the information from the approximate correlator, mediated by
the dispersion relation, only to determine the value of
parameters by a fit. This approach has been quite successful at
zero temperature and density, thanks to the fortuitous situation
that we roughly know how to parameterize the SF's in many cases.
Two important examples are the QCD sum rules pioneered by
Shifman, Vainshtein and Zakharov~\cite{SVZ,QCDSRrev}, and the
analysis of lattice QCD data~\cite{chu93}. So far, standard
parameterizations have included poles plus perturbative
continuum~\cite{shuryak}.

The success of such approaches heavily rests on our good
understanding of the qualitative behavior of SF's at zero $(T,\mu)$.
We can find other such favorable examples in the low-$T$
regime~\cite{hatsuda1}, where the shape of the SF has the same
qualitatively features of the zero $(T,\mu)$ case, or even in the
high-$T$ regime for simple models~\cite{huang93}, for which the
functional form of the SF is known.

The QCD sum rules approach has been extended also to systems
at finite temperature~\cite{Bochkarev86,Dosch88,hatsuda1}.
The lack of experimental data, and of
reliable nonperturbative calculations has prompted people to use
the same kind of parameterizations that have worked so well at
zero temperature with, at most, perturbative corrections. We
believe that physical results at finite $(T,\mu)$ can be strongly
biased by this assumption. In fact, recent interpretations of
lattice simulation data~\cite{hashimoto,karsch,chu94}
appear to indicate the existence of such problems.

The purpose of this work is to derive exact sum rules
that constrain the variation of SF's with $(T,\mu)$. In addition, we
apply these sum rules to the chiral phase transition, and demonstrate
that SF's in some channels are drastically modified compared to both
their zero $(T,\mu)$ and their perturbative shapes. This result
confirm our worries about non-trivial effect of finite $T$ or
baryon density on the shape of the SF.

Our derivation of these exact sum rules, based on the OPE
and the RGE, has a closer relation with the derivation of sum rules
for deep inelastic scatterings than with the QCD sum rule approach
of SVZ~\cite{SVZ}. In fact, we establish relationships between
moments of the SF and corresponding condensates as functions of
$(T,\mu)$, without assuming any functional form of the SF.

In the derivation process, we find that the logarithmic corrections
are essential to establish the exact sum rules. In contrast, the QCD
logarithmic corrections are only marginally relevant in the finite
energy sum rules, and hence are rarely discussed in the literature.
To properly take into account the logarithmic corrections, a repeated
partial integration method is used to match the relevant asymptotic
expansions.

  Since no further assumptions other than the validity
of the OPE and RGE are involved in the derivation, our
sum rules are very general and can be applied anywhere in the
$(T,\mu)$-plane, even near or at the boundary of a phase transition.

  The paper is organized as follows. In section II
we present the general derivation of exact sum rules in asymptotically
free theories. The matching of asymptotic behaviors of the correlator
and the dispersion integral, including explicitly their logarithmic
corrections, are carefully discussed.
In section III we illustrate each single step of the derivation
in a soluble model, the Gross-Neveu model in the large-$N$ limit.
In this model we can calculate exactly  all the relevant quantities
(spectral functions, Wilson coefficients, condensates, anomalous
dimensions and correlators in space-like region, etc.), and,
therefore, give a concrete example of how our method works.
The application of our method to the derivation of exact
sum rules for the mesonic channels in QCD is presented in section IV.
In the same section, we also discuss the
phenomenological consequences of the exact sum rules near
chiral restoration phase transitions.
Finally, in section V we summarize our work, draw some conclusions,
and discuss possible future directions.

\section{General derivation}
\label{general}
We start this section with a short review of the linear response
theory, the OPE and the RGE. Next we introduce a convenient
subtraction in the dispersion relation
for studying the dependence of SF's on $T$ and $\mu$.

Then, we present in detail a crucial part of our method:
how to match the asymptotic OPE expansion with a corresponding
asymptotic expansion of the SF and its dispersion
integral. This approach is necessary for properly taking into
account the logarithmic corrections, and studying the convergence
properties of the relevant moments of the SF. More naive approaches
not only yield, in general, incorrect sum rules, but might also fail
to recognize that a given sum rule does not exist in the first place,
since the integral of the SF involved is divergent.

Finally, a comparison of the two asymptotic expansions leads
to the desired exact sum rules. We end this section with some general
comments on the derivation and meaning of these sum rules.

\subsection{Linear Response}

The real-time linear response~\cite{negele} to an external source,
$S(x)$, coupled to a renormalized current $J(x)$ in the form of
$\Delta{\cal L}=S(x)J(x)$ is given by the retarded correlator:
\begin{equation}
K(x;T,\mu)\equiv
\theta(x_0)\langle [J(x),J(0)]\rangle_{T,\mu}\, ,
\label{corr}
\end{equation}
where the average is on the grand canonical ensemble specified by
temperature $T$ and chemical potential $\mu$. Disregarding possible
subtraction terms, which are $(T,\mu)$-independent since they are
related to short distance properties of the theory, we can write the
following dispersion relation
for the frequency dependence of the retarded correlation function:
\begin{equation}
\tilde{K}(\omega,\bbox{k};T,\mu)=\int_{0}^\infty du^2 \,
{\rho(u,\bbox{k};T,\mu)\over u^2-(\omega+i\epsilon)^2} \, .
\label{disp}
\end{equation}
For convenience, we discuss only the uniform limit
($\bbox{k}=0$) and, from now on, drop the momentum label.
We wish to emphasize that, even if
the derivation of the sum rules requires the analytic
continuation of Eq.~(\ref{disp}) into the Euclidean region
($\omega\rightarrow iQ$), the dispersion integral is, nevertheless,
determined by the singularities of $\rho(u;T,\mu)$
in the Minkowski region. One should keep clearly in mind the
distinction between these real-time singularities with the ones
related to screening phenomena.

\subsection{The OPE and the RGE}
In asymptotically free theories, the correlator
$\tilde{K}(\omega=iQ;T,\mu)$ is calculable in terms of the
OPE~\cite{brown} asymptotic series in the limit of
$Q^2\rightarrow\infty$. For a given
renormalization prescription, the asymptotic expansion reads
\begin{equation}
\label{ope}
\tilde{K}(iQ,\kappa;T,\mu)
\sim \tilde{K}_0(iQ,\kappa)
+ \sum_n C_n(Q^2,m^2(\kappa),g^2(\kappa),\kappa)
\langle [O_n]_\kappa\rangle_{T,\mu} \, ,
\end{equation}
where $m(\kappa)$, $g(\kappa)$, the $[O_n]_\kappa$'s and the $C_n$'s
are, respectively, the current quark mass, the coupling constant,
the renormalized composite operators, and their corresponding Wilson
coefficients at the subtraction scale $\kappa$. Here and in the following,
we use the symbol $\sim$ to denote asymptotic equality.
We explicitly singled out the perturbative term $\tilde{K}_0(iQ,\kappa)$,
which corresponds to the identity operator and, being independent of
$T$ and $\mu$, is calculated at $(T,\mu)=0$. This term is the only one
that is not suppressed by powers of $1/Q^2$ and can depend on an overall
subtraction scale, which we identify, for simplicity, with the same
subtraction scale $\kappa$ used in the renormalization of the operators.
It is important to note that the information of the
ensemble average is encoded in the matrix elements of composite
operators, while the Wilson coefficients, including $\tilde{K}_0$, are
independent of $T$ and $\mu$. Although the matrix elements
$\langle [O_n]_\kappa\rangle_{T,\mu}$ cannot be determined perturbatively,
the $Q^2$-dependence of the Wilson coefficient $C_n$ is controlled
by the following renormalization group equation~\cite{brown}
\begin{mathletters}
\begin{equation}
\Bigl[\kappa{d\over d\kappa} +2\Gamma_J-\Gamma_n\Bigr]
C_n(Q^2,m^2(\kappa),g^2(\kappa),\kappa)=0 \, ,
\end{equation}
where
\begin{equation}
\kappa\frac{d}{d\kappa}=
\kappa{\partial\over\partial\kappa}
+\beta{\partial\over\partial g}
-\Gamma_m{\partial\over\partial m} \, ,
\end{equation}
\label{rge}
\end{mathletters}
and $\Gamma_J$, $\Gamma_n$ and $\Gamma_m$ are the anomalous
dimensions for the current $J$, the operator $O_n$ and the
current quark mass $m$ respectively. For the purpose of illustration, we
are only considering operators that do not mix under renormalization, but
the mixing will be properly taken into account when necessary. The standard
approach~\cite{brown} to the renormalization group equation is the
introduction of a running coupling $g^2(Q)$ and a running mass $m(Q)$.
In asymptotically free theories $g^2(Q)$ vanishes logarithmically at large
$Q^2$. It is therefore meaningful to consider, in this limit, a perturbative
expansion of the  renormalization group equation functions
\begin{equation}
\Gamma_i=-\gamma_i g^2+{\cal O}(g^4)\, ,\,\,\, \text{and}
\,\,\, \beta=-b g^4+{\cal O}(g^6) \, ,
\label{rgef}
\end{equation}
where $i=J$, $n$, $m$, while $\gamma_i$, $b$ are pure numbers determined by
a one-loop calculation.
Within this perturbative context, Eq.~(\ref{rge}) can be solved~\cite{brown}:
\begin{equation}
C_n(Q^2,m(\kappa),g^2(\kappa),\kappa)=
\biggl({1\over Q^2}\biggr)^{\frac{d_n}{2}}
\biggl({g^2(Q)\over g^2(\kappa)}\biggr)^{\frac{2\gamma_J-\gamma_n}{2b}}
\times c_n\bigl(\frac{m^2(Q)}{Q^2},g^2(Q)\bigl) \, ,
\label{cn}
\end{equation}
where $d_n$ is the canonical dimension of the
composite operator $O_n$ minus the dimension of
$\tilde{K}$ in units of a mass, and the $c_n(m^2(Q)/Q^2,g^2(Q))$'s are
calculable perturbatively. Corrections to this result coming from
nonleading powers of $g^2$ in the expansions of the
renormalization group equation functions, see Eq.~(\ref{rgef}),
are analytic in $g^2(Q)=1/(b\ln Q^2)$ and hence can be absorbed
into $c_n$. Therefore, the leading
term in the large-$Q^2$ limit is totally characterized by the
one-loop structure of the theory.

\subsection{Subtraction and $(T,\mu)$-dependence}
  To study the temperature and chemical potential dependence of $\tilde{K}$ ,
we only need to consider the difference $\Delta\tilde{K}(iQ)$:
\begin{equation}
\Delta\tilde{K}(iQ)
\equiv \tilde{K}(iQ,\kappa;T,\mu)-\tilde{K}(iQ,\kappa;T',\mu')
= \int_0^\infty du^2\, {\Delta\rho(u)\over u^2+Q^2}\, ,
\label{ddisp}
\end{equation}
where $\Delta\rho(u)\equiv\rho(u;T,\mu)-\rho(u;T',\mu')$.
This subtraction is crucial to remove
$\tilde{K}_0(iQ,\kappa)$, which contains terms not suppressed by powers of
$1/Q^2$ and is explicitly dependent on the renormalization point.
The OPE asymptotic expansion of $\Delta\tilde{K}(iQ)$ is then
\begin{equation}
\Delta\tilde{K}(iQ)\sim
\sum_n \biggl({1\over Q^2}\biggr)^{\frac{d_n}{2}}
\biggl({g^2(Q)\over g^2(\kappa)}\biggr)^{\frac{2\gamma_J-\gamma_n}{2b}}
\times c_n(\frac{m^2(Q)}{Q^2},g^2(Q))
\Delta\langle [O_n]_\kappa\rangle \,
\label{dope}
\end{equation}
where $\Delta\langle[O_n]_\kappa\rangle$ denotes the difference
of the expectation values of $[O_n]_\kappa$ in the ensembles
specified by $(T,\mu)$ and $(T',\mu')$, respectively. Since the
current quark mass only runs logarithmically in asymptotically
free theories, we can safely ignore the corrections to the $c_n$'s
due to their dependence on $m^2(Q)/Q^2$ when in deep Euclidean limit.
Notice that the subtraction has made
$\Delta\tilde{K}(iQ)$ independent of the renormalization
point $\kappa$, i.e. $\Delta\tilde{K}(iQ)$ satisfies a homogeneous RGE.
\subsection{Asymptotic Expansion}
Since $c_n(m^2(Q)/Q^2,g^2(Q))$ is perturbative and hence can be
expanded in power of $g^2(Q)$ (and denoting the first non-vanishing
power as $\nu_n^{(0)}$), the left-hand side of Eq.~(\ref{ddisp})
can be expressed as a double (in $1/Q^2$ and $g^2(Q)$) asymptotic
expansion of the form:
\begin{equation}
\Delta\tilde{K}(iQ)\sim
\sum_{n,\nu = 0}^{\infty}
\frac{c_n^{(\nu)}(\kappa)\Delta\langle [O_n]_\kappa\rangle}{Q^{d_n}}
[g^2(Q)]^{\nu+\eta_n}\, ,
\label{asyope}
\end{equation}
where the exponents $\eta_n=\nu_n^{(0)}+(2\gamma_J-\gamma_n)/(2b)$
and the $Q^2$-independent coefficients $c_n^{(\nu)}(\kappa)$ are
again known perturbatively. It is important to emphasize here that
the leading term in $g^2(Q)$ for a given power of $1/Q^2$ is
controlled by the relevant anomalous dimensions.

We proceed by making an analogous asymptotic expansion of
$\Delta\rho(u)$ valid for $u^2\in(\lambda^2,\infty)$  with
$\lambda^2$ suitably large but otherwise arbitrary:
\begin{mathletters}
\begin{equation}
\Delta\rho(u)
\equiv \Delta\rho_{\text{power}}(u) + \Delta\rho_{\text{exp}}(u)
\end{equation}
with
\begin{equation}
\Delta\rho_{\text{power}}(u)
\sim \sum_{n=0}^\infty \frac{[\ln(u^2)]^{-\xi_n}}{u^{2(n+1)}}
     \sum_{\nu=0}^\infty a_{n}^{(\nu)}(T,\mu) [\ln(u^2)]^{-\nu} \, ,
\label{rholn}
\end{equation}
\end{mathletters}
where we have explicitly isolated in $\Delta\rho_{\text{exp}}(u)$
all terms that vanish exponentially when $u^2\rightarrow\infty$, such as
the pole contributions to $\Delta\rho(u)$ or terms containing the factor
$\exp(-u/T)$. To simplify the notation we have chosen the units such
that the running coupling has the form at one-loop level
$g^2(u)=1/(b\ln u^2)$ (or equivalently, energy scales are measured
in units of the relevant $\Lambda$-parameter).

This ansatz is sufficient to produce
an asymptotic series of the form of Eq.~(\ref{asyope}) to one-loop
level. More generally, one could
replace $1/\ln{u^2}$ with the full running coupling
$g^2(u)$ in the asymptotic sequence and generalize the method we are
going to describe; the general strategy involved in this generalization
can be found, for example, in Ref.~\cite{byz}. Since we are presently
only interested in one-loop calculations, we can regard $g^2(Q)$
and $1/\ln{Q^2}$ to be proportional.
It is easy to recognize that the sum over ``$n$'' is meant to match the
sum over mass dimension in Eq.~(\ref{asyope}), while the sum over
``$\nu$'' will match the sum over the order in perturbation series.
The existence of anomalous dimensions in the OPE makes it necessary
to introduce $\xi_n$ in the expansion for the spectral function.
In the following we ignore the dependence of $\xi_n$ on ``$n$'', as
we have already ignored the fact that there generally exist more than one
operator at a given dimension, and write $\xi$ to avoid a too cumbersome
notation. The complete notation will be restored when necessary.

  Before we proceed further, we wish to discuss whether it is possible
that additional terms might appear in
the expansion of $\Delta\rho_{\text{power}}(u)$ in Eq.~(\ref{rholn}).
In general, our asymptotic expansion procedure is
powerful enough to exclude this possibility within the framework of
the OPE. In fact, it allows to verify unambiguously that terms different
from the ones already present generate, when substituted
in the dispersion integral, terms that are missing in the OPE.
Let us examine two specific examples that might be suspected to exist
otherwise. First, dimensional arguments could suggest terms like
$g^2T/\sqrt{u^2}$. Our procedure shows that any $1/\sqrt{u^2}$ term
in $\Delta\rho(u)$ would
generate a $1/\sqrt{Q^2}$ term in $\Delta\tilde{K}(iQ)$, which does not
correspond to any known condensate, and it is therefore excluded.
The second example is given by terms such as $\ln(T^2/u^2)$, which are
naturally produced by elementary perturbative calculations of the
spectral function: are such terms present in Eq.~(\ref{rholn})?
It is indeed true that such terms appear in the
high-$T$ expansion of the spectral function at fixed $u^2$.
However, the expansion of the spectral function that is relevant for
comparing to the OPE is a high-$u^2$ expansion at fixed $T$.
In the next section an explicit calculation in the
Gross-Neveu model will illustrate the general fact that, contrary to the
high-$T$ expansion, the high-$u^2$ expansion does not generate terms like
$\ln(T^2/u^2)$.

At this stage, we can already recognize a fundamental, and often
overlooked, characteristic of the spectral function. If we insert
the term $\Delta\rho_{\text{exp}}(u)$ in the dispersion integral, we only
obtain pure powers of $1/Q^2$, since the exponential convergence
allows a naive expansion of the factor $1/(u^2+Q^2)$
(Watson's Lemma~\cite{bleistein}). Therefore
logarithmic corrections, i.e. powers of $g^2(Q)$, come solely from the
$\Delta\rho_{\text{power}}(u)$ term.
Because we know that the running coupling $g^2(Q)$ is always present in
the OPE series, the term $\Delta\rho_{\text{power}}(u)$ must be present
in the subtracted SF, and it obviously dominates the asymptotic regime
($u^2\rightarrow\infty$). This fact immediately implies that only a
finite number of moments of the subtracted SF can possibly be finite,
i.e. the naive expansion of the factor $1/(u^2+Q^2)$ is generally wrong,
and that logarithmic corrections play a important role.

A standard method to tackle the dispersion integral in the large-$Q^2$
limit is the Mellin transform. However, the use of Mellin transform
methods~\cite{bleistein} is extremely cumbersome when logarithms
appear in the denominator. Since inverse logarithms cannot be avoided
in the spectral function,  we need to resort to other means.

We carry out the dispersion integral of the $\Delta\rho_{\text{power}}(u)$
term by splitting the integral in Eq.~(\ref{ddisp}) into three intervals:
$(0,\lambda^2)$, $(\lambda^2,Q^2)$ and $(Q^2,\infty)$.

The integral over the first interval can be naively expanded in powers of
$1/Q^2$, since there are no convergence problems:
\begin{equation}
\Delta\tilde{K}_1(iQ)
\equiv \int_0^{\lambda^2}du^2\,
{\Delta\rho_{\text{power}}(u)\over u^2+Q^2}
\sim {1\over Q^2} \sum_{n=0}^\infty{(-1)^n\over Q^{2n}}
\int_0^{\lambda^2}du^2\, u^{2n}\Delta\rho_{\text{power}}(u)\, .
\end{equation}
In particular, the leading term in $1/Q^2$ is
\begin{equation}
\Delta\tilde{K}_1(iQ)
\sim \frac{1}{Q^2}
\int_0^{\lambda^2}du^2\, \Delta\rho_{\text{power}}(u) + {\cal O}(1/Q^4)\, .
\label{int1asym}
\end{equation}

In the second interval, we use the asymptotic  form of
$\Delta\rho_{\text{power}}$ given in Eq.~(\ref{rholn}) and obtain (for
instance by repeatedly integrating by parts)
\begin{eqnarray}
\Delta\tilde{K}_2(iQ)
&\equiv&\int_{\lambda^2}^{Q^2} du^2\,
         {\Delta\rho_{\text{power}}(u)\over u^2+Q^2}
\sim{1\over Q^2}\sum_{n,\nu=0}^\infty a_n^{(\nu)} \biggl\{
 \frac{(-1)^n}{Q^{2n}}
\frac{(\ln Q^2)^{\alpha}-(\ln \lambda^2)^{\alpha}}{\alpha} \nonumber\\
&+& \sum_{l=0\atop l\neq n}^\infty \frac{(-1)^l}{Q^{2l}}
\frac{\Gamma(\alpha,(n-l)\ln\lambda^2)-\Gamma(\alpha,(n-l)\ln Q^2)}
     {(n-l)^{\alpha}} \biggr\}
\label{int2}
\end{eqnarray}
where $\alpha\equiv 1-\nu-\xi$, and $\Gamma[\alpha,z]$ is the incomplete
Gamma function.
Notice that when $\alpha\equiv1-\nu-\xi=0$ the correct result for the term
with $l=n$ is  $\ln\frac{\ln Q^2}{\ln\lambda^2}$, which also corresponds
to its limiting value:
$\lim_{\alpha\to 0} [(\ln Q^2)^{\alpha}-(\ln \lambda^2)^{\alpha}]/\alpha
=\ln\frac{\ln Q^2}{\ln\lambda^2}$.
We then substitute in the second incomplete Gamma function that appears in
Eq.~(\ref{int2}) the following asymptotic expression which is valid for
$|z|\to \infty$ and $-3\pi/2< \text{arg} z <3\pi/2$:
\begin{equation}
\Gamma[\alpha,z]\equiv\int_z^\infty dx\, x^{\alpha-1} e^{-x}
\sim\, z^{\alpha-1} e^{-z}\sum_{m=0}^\infty
{(-1)^m\Gamma(1-\alpha+m)\over z^m\Gamma(1-\alpha)}\, .
\label{asygamma}
\end{equation}
The resulting expression for $\Delta\tilde{K}_2(iQ)$ is
\begin{eqnarray}
\Delta\tilde{K}_2(iQ)
&\sim&{1\over Q^2} \sum_{l,\nu=0}^\infty\frac{(-1)^l}{Q^{2l}} \biggl[
\sum_{n=0\atop n\neq l}^\infty a_n^{(\nu)}
\frac{\Gamma(\alpha,(n-l)\ln\lambda^2)}{(n-l)^{\alpha}}
- a_l^{(\nu)} \frac{(\ln \lambda^2)^{\alpha}}{\alpha} \biggr]
\label{dk2asym}
                                                           \\
&-&{1\over Q^2}\sum_{n,\nu=0}^\infty {a_n^{(\nu)}\over Q^{2n}}\biggl[
\sum_{l=0\atop l\neq n}^\infty \sum_{m=0}^\infty \frac{(-1)^{l+m}}
 {(n-l)^{m+1}[\ln Q^2]^{m+1-\alpha}}
\frac{\Gamma(m+1-\alpha)}{\Gamma(1-\alpha)}-
\frac{(-1)^n [\ln Q^2]^{\alpha}}{\alpha} \biggr]\, , \nonumber
\end{eqnarray}
where again $(\ln\lambda^2)^{\alpha}/\alpha$ and
$(\ln Q^2)^{\alpha}/\alpha$ should be understood as
$\ln(\ln\lambda^2)$ and $\ln(\ln Q^2)$ when
$\alpha\equiv 1-\nu-\xi=0$.

Let us notice that in the first line of Eq.~(\ref{dk2asym}) any term with
$l$ such that the integral
$\int_{\lambda^2}^\infty du^2\,u^{2l}\Delta\rho_{\text{power}}(u)$ is
finite can be formally identified as
\begin{equation}
\sum_{\nu=0}^\infty\biggl[\sum_{n=0\atop n\neq l}^\infty a_n^{(\nu)}
\frac{\Gamma(\alpha,(n-l)\ln\lambda^2)}{(n-l)^{\alpha}}
- a_l^{(\nu)} \frac{(\ln \lambda^2)^{\alpha}}{\alpha}\biggr]=
\int_{\lambda^2}^\infty du^2\,u^{2l}\Delta\rho_{\text{power}}(u)\, .
\label{formom}
\end{equation}
In particular, the leading $1/Q^2$ term in Eq.~(\ref{dk2asym}) is
\begin{eqnarray}
\Delta\tilde{K}_2(iQ)
&\sim&{1\over Q^2} \sum_{\nu=0}^\infty\biggl[
\sum_{n=1}^\infty a_n^{(\nu)}
\frac{\Gamma(\alpha, n\ln\lambda^2)}{n^{\alpha}}
- a_0^{(\nu)} \frac{(\ln \lambda^2)^{\alpha}}{\alpha} \biggr]\\
&+&{1\over Q^2}\sum_{\nu=0}^\infty a_0^{(\nu)}\biggl[
\sum_{l=1}^\infty \sum_{m=0}^\infty \frac{(-1)^{l}}
 { l^{m+1}[\ln Q^2]^{m+1-\alpha}}
\frac{\Gamma(m+1-\alpha)}{\Gamma(1-\alpha)}+
\frac{ [\ln Q^2]^{\alpha}}{\alpha} \biggr]
\label{int2asym}
 + {\cal O}\left(\frac{1}{Q^4}\right) \, . \nonumber
\end{eqnarray}

We can similarly expand the integral over the third interval
\begin{eqnarray}
\Delta\tilde{K}_3(iQ)
&\equiv&\int_{Q^2}^\infty
du^2\, {\Delta\rho_{\text{power}}(u)\over u^2+Q^2}
\sim -{1\over Q^2}
\sum_{n,\nu=0}^\infty a_n^{(\nu)}\sum_{l=1}^\infty (-Q^2)^l
\frac{\Gamma(\alpha,(n+l) \ln Q^2)}{(n+l)^{\alpha}}\nonumber \\
&\sim&\mbox{}-{1\over Q^2} \sum_{n,\nu=0}^\infty{a_n^{(\nu)}\over Q^{2n}}
\sum_{l=1}^\infty \sum_{m=0}^\infty
\frac{(-1)^{l+m}}{(n+l)^{m+1} [\ln Q^2]^{m+1-\alpha}}
{\Gamma(m+1-\alpha)\over\Gamma(1-\alpha)} \nonumber \\
&\sim&\mbox{}-{1\over Q^2} \sum_{\nu=0}^\infty a_0^{(\nu)}
\sum_{l=1}^\infty \sum_{m=0}^\infty
\frac{(-1)^{l+m}}{ l^{m+1} [\ln Q^2]^{m+1-\alpha}}
{\Gamma(m+1-\alpha)\over\Gamma(1-\alpha)}
+ {\cal O}\left(\frac{1}{Q^4}\right) \, .
\label{int3}
\end{eqnarray}

  In the end we add the leading $1/Q^2$ contributions from
Eqs.~(\ref{int1asym}), (\ref{int2asym}) and (\ref{int3}) to the
corresponding contribution
from the naive expansion of $\Delta\rho_{\text{exp}}$ and obtain
\begin{eqnarray}
\Delta\tilde{K}(iQ)
&\equiv&\int_{0}^\infty
du^2\, {\Delta\rho(u)\over u^2+Q^2}
\sim \frac{\overline{\Delta\rho}}{Q^2}
+\frac{[\ln Q^2]^{1-\xi}}{Q^2}
\sum_{\nu=0}^\infty \frac{a_0^{(\nu)}}{[\ln Q^2]^{\nu}}\biggl[
\frac{1}{1-\xi-\nu} \nonumber \\ & &\,\,\,\,\,
+2\sum_{l=1\atop m=0}^\infty\frac{(-1)^l }{ [l\, \ln Q^2]^{2m+2}}
\frac{\Gamma(2m+1+\nu+\xi)}{\Gamma(\nu+\xi)}\biggr]
+ {\cal O}\left(\frac{1}{Q^4}\right)\, ,
\label{leadqrho}
\end{eqnarray}
where $\overline{\Delta\rho}$ is defined by
\begin{eqnarray}
\overline{\Delta\rho}&\equiv&
\int_0^{\lambda^2} du^2\, \Delta\rho_{\text{power}}(u)
+ \int_0^\infty du^2\,\Delta\rho_{\text{exp}}(u) \nonumber\\
&+&\sum_{n=1\atop\nu=0}^\infty a_n^{(\nu)}\,
\frac{\Gamma(1-\nu-\xi,n\ln\lambda^2)}{ n^{1-\nu-\xi}}
-\sum_{\nu=0}^\infty  a_0^{(\nu)}
\frac{(\ln \lambda^2)^{1-\nu-\xi}}{1-\nu-\xi}\, .
\label{drhobar}
\end{eqnarray}
Since the integral $\int^{\infty} dx x^{-1} (\ln x)^{-\xi}$ can be
finite only if $\xi>1$, we can use Eq.~(\ref{formom}) to identify
$\overline{\Delta\rho}$ as the zeroth moment of the subtracted spectral
function, only when the asymptotic expansion of the subtracted SF in
Eq.~(\ref{rholn}) yields $\xi>1$. Otherwise, if $\xi\leq 1$, the zeroth
moment is infinite, though $\overline{\Delta\rho}$ as defined by
Eq.~(\ref{drhobar}) still exists.
\subsection{Exact Sum Rules}
We derive the sum rules by comparing the
coefficient of $1/Q^2$ in Eq.~(\ref{leadqrho})
\begin{equation}
Q^2 \int_{0}^\infty du^2\, {\Delta\rho(u)\over u^2+Q^2}
\sim \overline{\Delta\rho}\,
+ a_0^{(0)}\times\left\{\begin{array}{c}
  \frac{\textstyle [\ln Q^2]^{1-\xi} }{\textstyle (1-\xi)} \\[0.6cm]
  \ln(\ln Q^2)   \end{array}
 + {\cal O}\left({1\over Q^2},{1\over[\ln Q^2]^\xi}\right)
   \quad
    \begin{array}{l}
    \mbox{if $\xi\neq 1$} \\[0.6cm]
    \mbox{if $\xi=1$}
    \end{array}\right.
\label{leadingrh}
\end{equation}
and the corresponding coefficient in Eqs.~(\ref{asyope})
\begin{equation}
Q^2 \Delta\tilde{K}(iQ)
\sim c_1^{(0)}(\kappa)\Delta\langle [O_1]_\kappa\rangle
[g^2(Q)]^{\eta}
+{\cal O}\left(\frac{1}{Q^2},[g^2(Q)]^{\eta+1}\right) \, ,
\end{equation}
where we are only considering cases with $n=1$ and $d_1=2$.
Furthermore, we are presently interested in exact sum rules that
can be derived with one-loop calculations and, therefore, we only compare
leading orders in $1/\ln{Q^2}$. Then there exist three possibilities,
depending on the value of $\eta$, which can be calculated using the OPE.

\noindent (1) If $\eta=0$, then $\xi=2$ (or a bigger integer) and
\begin{equation}
\overline{\Delta\rho}=\int_0^\infty du^2\, \Delta\rho(u) =
c_1^{(0)}(\kappa)\Delta\langle[O_1]_\kappa\rangle \, .
\label{cons}
\end{equation}
This is the result one would get by naively expanding $Q^2/(u^2+Q^2)$
without worrying about the convergence of the moments and disregarding
logarithmic corrections in the OPE.

\noindent (2) If $\eta>0$, then $\xi=1+\eta$ and
\begin{equation}
\overline{\Delta\rho}=\int_0^\infty du^2\, \Delta\rho(u) = 0 \, .
\label{nonc}
\end{equation}

\noindent (3) If $\eta<0$ (positive powers of $\ln Q^2$ in Eq.~(\ref{dope}))
 or if the term $\ln\ln Q^2$ appears in Eq.~(\ref{dope}), then
 $\xi<1$ or $\xi=1$, respectively. This implies that
\begin{equation}
\overline{\Delta\rho}\neq\int_0^\infty du^2\, \Delta\rho(u) = \infty \, .
\end{equation}
We remark that, even in this case when the moment is not finite,
the asymptotic expansion is still well defined. It is nice to see
that whether the zeroth moment of the subtracted SF exists is reflected
directly through the leading power of $g^2(Q)$ in the OPE series.

Our main results, Eq.~(\ref{cons}) and Eq.~(\ref{nonc}), can be
expressed in physical terms as follows. The zeroth moment of a SF for
a current $J$ whose OPE series yields $\eta > 0$ is independent
of $T$ and $\mu$, while the same moment for a current with $\eta=0$
changes with $T$ and $\mu$ proportionally to the corresponding
change(s) of the condensate(s) of the leading operator(s). Although
$c_1^{(0)}(\kappa)$ and $\Delta\langle[Q_1]_\kappa\rangle$ can
separately depend on $\kappa$, their product must be independent
of $\kappa$, since the zeroth moment is independent of $\kappa$.

\subsection{Discussion}
At this point several general comments are appropriate:

\noindent (1)
Our derivation relies on the fact that an asymptotically free theory
allows a perturbative expansion at short distances, making practical the
use of the OPE and of the RGE. We understand why only short distance
physics is involved if we realize that the integral over frequencies
reduces Eq.~(\ref{corr}) to the ensemble average of the equal-time
commutator of the currents. Therefore, results such as Eq.~(\ref{cons})
and Eq.~(\ref{nonc}) are completely determined by the one-loop structure
of the theory and the particular current under exam.

\noindent (2)
Flavor, or other non-dynamical quantum numbers, does not change the
expansions at the one-loop level in an essential way. Therefore,
one can derive analogous sum rules by using other kind of subtractions,
instead of the one we adopted. One such example is given by the exact
Weinberg sum rules $(T=0)$ in the chiral limit~\cite{weinberg}.

\noindent (3)
The derivation of sum rules for higher moments of the SF requires the
complete cancelation of all the lower dimensional operator terms, not
just the leading $g^2(Q)$. In particular, one also needs current
quark mass corrections to the Wilson coefficients. Without appropriate
subtractions, higher moments do not even converge~\cite{kapusta}.

\noindent (4)
It is essential to properly take into account the logarithmic
corrections when deriving exact sum rules, since the logarithmic
corrections not only dictate whether $\overline{\Delta\rho}$ satisfy
Eq.~(\ref{cons}) or Eq.~(\ref{nonc}) but they also control the very
existence of $\overline{\Delta\rho}$~\cite{kapusta}.
This procedure is in sharp contrast with the usual QCD sum rule approach,
where the convergence issue is by-passed by applying the Borel improvement
by explicitly introducing a cut-off parameter (the Borel mass).

\noindent (5)
We believe that the $(T,\mu)$-dependent part of the leading condensate
appearing in Eq.~(\ref{cons}) does not suffer from the infrared
renormalon ambiguity. In fact, only the perturbative term $\tilde{K}_0$
can generate contributions to the leading condensate that are dependent
on the prescription used to regularize these renormalons.
But $\tilde{K}_0$ is independent of $T$ and $\mu$:
any prescription dependence cancels out when we make the subtraction
in Eq.~(\ref{ddisp}). On the contrary, unless we generalize
Eq.~(\ref{ddisp}) and make other subtractions, sum rules that involve
non-leading condensates are, in principle, ambiguous.

\noindent (6)
It is well-known that conserved operators are not renormalized and,
barring anomalous violations, verify the same ``classical'' identities
that can be derived at the tree level. This fact is also verified in
the sum rules. In fact when both the currents and the operators are
conserved ($\eta=0$) and we obtain the result of Eq.~(\ref{cons}), i.e.
one can use the ``naive asymptotic expansion'' to derive the sum rule.
Conservation of the current alone is not enough to warrant a
``classical'' identity.

\section{Gross-Neveu model}
\label{GNmodel}
In the preceding section we have derived sum rules valid for any
asymptotically free theory. In this section we illustrate the
procedure in the 1+1 dimensional Gross-Neveu model~\cite{gn}
in the infinite-$N$ limit. On one hand, we can derive the sum rules
in Eqs.~(\ref{cons}) and (\ref{nonc}) by explicitly
calculating Wilson coefficients, $\beta$- and  $\Gamma$-functions in
the vector and pseudoscalar channels, following the general procedure
discussed in section~II. On the other hand, since this model is soluble,
we can obtain the exact spectral function at arbitrary $(T,\mu)$, and then
explicitly verify both that the sum rules are satisfied and that the
asymptotic expansion of the spectral function has the form given in
Eq.~(\ref{rholn}). Moreover, we can also calculate the condensates, and
therefore explicitly check that the OPE really matches the asymptotic
expansion of the exact spectral integral.

The Lagrangian of the Gross-Neveu model is
\begin{equation}
{\cal L} = \bar{\psi}i{\gamma\cdot\partial}\psi+{g^2\over 2N}
\left[(\bar{\psi}\psi)^2 + (\bar{\psi}i\gamma_5\psi)^2\right]\, ,
\end{equation}
where $\psi$ is a two-component Dirac spinor and has $N$-component
in the internal space. Equivalently, we can write
\begin{equation}
{\cal L}=\bar{\psi}(i\gamma\cdot\partial-\sigma-i\gamma_5\pi)\psi
-{N\over 2g^2}(\sigma^2+\pi^2),
\end{equation}
where $\sigma$ and $\pi$ are auxiliary fields. The
coupling constant $g^2$ is independent of $N$ and held fixed
in the limit $N\to\infty$. This model is asymptotically free
when $D=1+1$, and the chiral symmetry is dynamically broken at
$T=0$ and $\mu=0$ to the leading order in $1/N$~\cite{gn}.
In the following we give the exact solution at
$N\rightarrow\infty$ with finite $(T,\mu)$.
Although formulas are explicitly considered in the symmetry breaking
phase, they are also valid in the symmetric phase
provided that the vanishing limit of certain condensates (such as the
dynamical fermion mass) is properly taken.

We wish to remark on the well-known fact that the limit $N\to\infty$
here should not be interpreted as a starting point for an expansion of
the model at finite $N$, but rather as different model in itself, which
is in fact the model we decided to use for the purpose of illustration.
Moreover, there exist arguments~\cite{wolff,gatto} suggesting that
this model in the limit $N\to\infty$ is actually more relevant to
$3+1$ phenomenology than the model with finite $N$.

\subsection{Exact solution in the large-$N$ limit}
  Since the procedure to obtain the exact solution in the
large-$N$ limit is rather standard and some of the intermediate
steps can be found in the literature, see for
instance Refs.~\cite{gn,ma,wolff,huang93,bernd},
we only give the definitions and final results.

\subsubsection{Gap equation and phase diagram}
  The gap equation and the phase diagram can be derived from
the first derivative of the effective potential
\begin{equation}
{\partial V(\sigma^2)\over\partial\sigma^2}
= {N\over 2\pi}
\Bigl\{{\pi\over g^2}
+ T\int dk
\sum_n{1\over (i\omega_n+\mu)^2-(k^2+\sigma^2)}\Bigr\}
\Bigg|_{\omega_n=(2n+1)\pi T}\, .
\end{equation}
We introduce a momentum cutoff $\Lambda$ and add the counterterm
${N\over 4\pi}\ln(4\Lambda^2/\kappa^2)$, where $\kappa$ is
the subtraction point.
Then the renormalized effective potential becomes
\begin{eqnarray}
{\partial V(\sigma^2)\over\partial\sigma^2}
&=&{N\over 2\pi} \Bigl\{\frac{\pi}{g^2} +
\frac{1}{2}\ln\frac{\sigma^2}{\kappa^2}+\int_0^\infty\!dk \,
\frac{f(\sqrt{k^2+\sigma^2};T,\mu)}{\sqrt{k^2+\sigma^2}} \Bigr\}
\nonumber \\ &=&{N\over 2\pi}
\Bigl\{\frac{1}{2}\ln\frac{\sigma^2}{M^2}
+\int_0^\infty\! dk\,
\frac{f(\sqrt{k^2+\sigma^2};T,\mu)}{\sqrt{k^2+\sigma^2}} \Bigr\} \, ,
\label{effpotp}
\end{eqnarray}
where we have defined
\begin{equation}
f(x;T,\mu)\equiv \frac{1}{e^{(x+\mu)/T}+1} +\frac{1}{e^{(x-\mu)/T}+1} \, ,
\end{equation}
and, in the second equality, we have eliminated the subtraction point by
introducing the dynamical fermion mass at $(T,\mu)=0$, i.e.
$M=\kappa \exp(-\pi/g^2)$.
The effective potential is then obtained by integrating Eq.~(\ref{effpotp}):
\begin{equation}
V(\sigma^2)
={N\over 4\pi} \left\{ \sigma^2\Bigl(\ln{\sigma^2\over M^2} -1\Bigr)
+4 T \int_0^\infty dk\,
\ln \left[   \frac{(1+e^{-(k+\mu)/T})(1+e^{-(k-\mu)/T}) }
 {(1+e^{-(\sqrt{k^2+\sigma^2}+\mu)/T})(1+e^{-(\sqrt{k^2+\sigma^2}-\mu)/T}) }
   \right]\right\}\, .
\label{effpot}
\end{equation}
The gap equation is simply given by
\begin{equation}
{\partial V\bigl(\sigma^2\bigr)\over \partial\sigma}
\Bigg|_{\sigma=m(T,\mu)} =0\, .
\label{gap_eq}
\end{equation}
The system possesses ``metastable states'' at those values of $(T,\mu)$
for which the gap equation has more than one minimum solutions:
the solution with lowest $V\bigl(m^2(T,\mu)\bigr)$ selects the true
ground state. When $(T,\mu)$'s are such that $m(T,\mu)=0$ is the only
minimum solution to the gap equation, these points in the
$(T,\mu)$-plane defines a second order phase transition line, which
separate the symmetry-broken phase, i.e. $m(T,\mu)\neq 0$, from the
symmetric phase, i.e. $m(T,\mu)=0$. Taking the limit $\sigma=0$ in
Eq.~(\ref{effpotp}), after some suitable rewriting, this critical line
obeys the following equation
\begin{equation}
\ln(2)
= {1\over 2}\int_0^M{dk\over k}
{\sinh(k/T)\over\cosh[(k+\mu)/2T]
                  \cosh[(k-\mu)/2T]}
- \int_M^\infty {dk\over k}\, f(k;T,\mu)\, ,
\label{gap0}
\end{equation}
whose numerical solutions are plotted in Fig.~\ref{fig1} in full line.
The full line turns into a dotted line at the heavy dot, where
another minimum (dash line) appears and becomes lower than the
minimum given by Eq.~(\ref{gap0}). Since the mass gap is finite on
the dash line (given by the solution of $V(\sigma^2)=V'(\sigma^2)=0$
with $\sigma\neq 0$, where prime denote a derivative with respect
to $\sigma^2$), the dash line is a first order phase boundary.
The critical temperature at $\mu=0$ is given by
$T_c=(e^{\gamma_E}/\pi)M\approx 0.566933M$. The ``tricritical point''
(the heavy dot in Fig.~\ref{fig1}) can be found by imposing that
$V'(\sigma^2\rightarrow 0)=V''(\sigma^2\rightarrow 0)=0$:
$T_3\approx 0.318328 M, \mu_3 \approx 0.608221 M$~\cite{wolff}.
When $T=0$ the chemical potential at which the first order phase
transition takes place is $\mu_c=M/\sqrt{2}$~\cite{wolff}.

\subsubsection{Bubble graph and correlator in space-like region}
  The basic ingredient to build up the correlator is the fermion
bubble graph. In the Euclidean region the pseudoscalar bubble, i.e.
the free bubble graph between two $J_5\equiv\bar{\psi}i\gamma_5\psi$
currents, can be conveniently expressed as
\begin{mathletters}
\begin{equation}
\Pi(\omega)=\Pi^{(0)}(\omega)+\Delta\Pi(\omega)\, ,
\end{equation}
where
\begin{equation}
\Pi^{(0)}(\omega)|_{\omega=iQ}=-{N\over\pi}
           \ln\bigl({\kappa\over m}\bigr)
\mbox{}+{N\over 2\pi}{1\over\sqrt{1+4m^2/Q^2}}\ln
\Bigl({\sqrt{Q^2+4m^2}+\sqrt{Q^2}\over\sqrt{Q^2+4m^2}-\sqrt{Q^2}}\Bigr)
\end{equation}
and
\begin{equation}
\Delta\Pi(\omega)|_{\omega=iQ}={N\over\pi}\int_0^\infty dk\,f(E_k;T,\mu)
{E_k\over E_k^2+Q^2/4}\, .
\end{equation}
\end{mathletters}
Here and in the following $\kappa$ is always the renormalization
subtraction point, $m=m(T,\mu)$ is the dynamically generated fermion mass,
and $E_k=\sqrt{m^2+k^2}$. We shall also use the short-hand notation
\begin{equation}
\langle\!\langle A\rangle\!\rangle\equiv
{1\over 2\pi}\int_{-\infty}^\infty {dk\over E_k}\, A(k) \,
 f(E_k;T,\mu) \, .
\end{equation}
For instance, the use of this notation makes possible to write the gap
equation as
\begin{equation}
\ln\frac{\kappa}{m}-\frac{\pi}{g^2}
= \int_0^\infty{dk\over E_k}\, f(E_k;T,\mu)
= \pi\langle\!\langle E_k^0\rangle\!\rangle.
\label{gapsh}
\end{equation}

  In the large-$N$ limit, the correlator in pseudoscalar channel is
given by the geometric sum of the bubble graph:
\begin{equation}
\langle J_5 J_5\rangle_{T,\mu}(\omega)=
\Pi(\omega)\left(1+{g^2\over N}\Pi(\omega)\right)^{-1}.
\end{equation}

In the Euclidean region, we can then expand this correlator in the
$Q^2\rightarrow\infty$ limit as
\begin{eqnarray}
{1\over N}\langle J_5 J_5\rangle_{T,\mu}(iQ)
&=& {\ln(Q^2/\kappa^2)/2\pi\over 1+g^2\ln(Q^2/\kappa^2)/2\pi}
-2g^2(Q){m^2/g^4\over Q^2}\Bigl(1-{g^2(Q)\over 2\pi}\Bigr) \nonumber\\
&+& \Bigl({g^2(Q)\over g^2}\Bigr)^2 {4\over Q^2}
\Bigl(\langle\!\langle E_k^2\rangle\!\rangle-{m^2\over 2}
\langle\!\langle E_k^0\rangle\!\rangle\Bigr)
+{\cal O}\Bigl({1\over Q^4}\Bigr)\, ,
\label{corrQ}
\end{eqnarray}
where we have introduced the running coupling constant
\begin{equation}
g^2(Q)\equiv{g^2\over 1+(g^2/2\pi)\ln(Q^2/\kappa^2)} \, ,
\end{equation}
and used the gap equation Eq.~(\ref{gapsh}).

We can recognize the first term in Eq.~(\ref{corrQ}) as the perturbative
contribution at $(T,\mu)=0$, $\langle J_5 J_5\rangle^{(0)}(iQ)$, which
corresponds to the term $\tilde{K}_0(iQ)$ in the generic OPE expansion
shown in section~\ref{general}. The other two terms in Eq.~(\ref{corrQ})
can be interpreted as contributions from the condensates. As we will
verify later, in fact,
$m^2/g^4 = \langle[(\bar{\psi}\psi)^2]\rangle_{T,\mu}$, while
$\langle\!\langle E_k^2\rangle\!\rangle
  -{1\over 2}m^2\langle\!\langle E_k^0\rangle\!\rangle
=\langle[\theta_{00}]\rangle_{T,\mu}$ is the condensate of the
energy-momentum tensor. We shall see that the perturbative contribution and
the coefficients of the condensates are indeed the ones obtained by the
OPE of $J_5J_5$, and therefore the expansion of the correlator shown in
Eq.~(\ref{corrQ}) is of the form of the OPE.

Note that in defining the condensates we have absorbed factors of $N$
in order to simplify the notation.

\subsubsection{Spectral function}
  To obtain the spectral function we need the bubble
graph in the time-like region $\omega^2>4m^2$:
\begin{equation}
\frac{1}{N}\text{Im}\bigl(\Pi(\omega)\bigr)
=\epsilon(\omega)
{\theta(\omega^2-4m^2)\over 4\sqrt{1-4m^2/\omega^2}}
\times \Bigl\{\tanh\Bigl({|\omega|/2+\mu\over 2T}\Bigr)
      +\tanh\Bigl({|\omega|/2-\mu\over 2T}\Bigr)\Bigr\}
\label{impi}
\end{equation}
and
\begin{equation}
 \frac{1}{N}\text{Re}\Bigl(\Pi(\omega)\Bigr)
=-{1\over g^2}
+\frac{\tanh^{-1}\sqrt{1-4m^2/\omega^2}}{\pi\sqrt{1-4m^2/\omega^2}}
+\frac{\omega^2}{4\pi} {\cal P} \!\int_0^\infty \frac{dk}{E_k}\,
\frac{f(E_k;T,\mu)}{E_k^2-\omega^2/4}\, ,
\label{repi}
\end{equation}
where ${\cal P}\!\int$ stands for the principle value of the integral.

 The complete spectral function has the form
\begin{equation}
\rho(\omega)=\pi\delta(\omega^2-m_\pi^2){N g_\pi^2(T)\over g^4}
+\theta(\omega^2-4m^2)\rho_{\text{cont}}(\omega)\, .
\label{rho_full}
\end{equation}
The pole contribution comes from the bound state (the pion), whose
mass is the solution to the equation
\begin{equation}
\frac{N}{g^2}+\Pi(\omega)\bigg|_{\omega=m_\pi}=0
\end{equation}
in the region $(0,2m)$. This equation has
in fact solution $m_\pi=0$, and the coupling constant of this massless pion
to its constituents  is given by
\begin{equation}
g_\pi^2(T)=\Bigl({\partial\over
\partial\omega^2}\Pi(\omega)/N\Bigr)^{-1}_{\omega^2=m_\pi^2=0}\, .
\end{equation}
The continuum part of the spectral function is related to the bubble
$\Pi(\omega)$ in the time-like region through
\begin{equation}
\rho_{\text{cont}}(\omega)={ \text{Im}\Pi(\omega)\over
(1+g^2\text{Re}\Pi(\omega)/N)^2+(g^2 \text{Im}\Pi(\omega)/N)^2}\, .
\end{equation}

  In order to get an idea of the dependence of the spectral function
on $(T,\mu)$, we plot $\rho(\omega)$ for several typical values of
$(T,\mu)$, which  are indicated in the phase diagram in Fig.~\ref{fig1}.
In Fig.~\ref{fig2} we show $\rho(\omega)$ for values of $(T,\mu)$ that
go from the symmetry-broken phase to the symmetric phase through the
second order line, while in Fig.~\ref{fig3} we show $\rho(\omega)$ as
$(T,\mu)$ go through the first order boundary.
We immediately notice that the
behavior of the spectral function changes drastically near the
second order phase boundary. This is a clear reflection of the
low energy critical phenomena related to the continuous phase transition.
In fact, the pseudoscalar correlator (in the symmetric phase only,
since it is then degenerate with the scalar correlator) at $Q^2=0$
can be regarded as the chiral susceptibility, and should diverge as
$|T-T_c|^{-\gamma}$ (with $\gamma=1$ in this model).
At the boundary of the first order phase transition (Fig.~\ref{fig3}),
but still relatively close to the ``tricritical point'',
the same qualitative deformation is present,
since a weak first order transition shares certain qualitative
features of a second order transition, even if we do not expect any
real divergence.

It is this kind of dramatic behavior of the spectral function near
phase transitions that makes practical parameterizations nearly
impossible. The chiral restoration in QCD is expected to be
either continuous, or at least a smooth crossover, and the
spectral functions in scalar and pseudoscalar channels should show
a behavior qualitatively similar to the Gross-Neveu model, i.e. strong
peak right above the origin near the phase transition. We shall comment
more on this issue after we derive the sum rules for QCD in
section~\ref{QCD}.

\subsubsection{Recovering the OPE series}
  It is instructive to expand $\rho(\omega)$ in
the limit $\omega^2\rightarrow\infty$. We shall see that this asymptotic
expansion is of the form assumed in section~\ref{general}. Furthermore,
once substituted in the dispersion integral, it generates the OPE series
of Eq.~(\ref{corrQ}) according to the procedure
developed in section~\ref{general}.

To the leading order in $1/\omega^2$, we find
\begin{equation}
{g^2\over N}\text{Im}\bigl(\Pi(\omega)\bigr)\sim
g^2\epsilon(\omega) \Bigl\{{1\over 2}+{m^2\over\omega^2}
+{\cal O}\bigl({1\over \omega^4},e^{-|\omega|/2T}\bigr)\Bigr\}
\end{equation}
and
\begin{equation}
1+{g^2\over N}\text{Re}\Bigl(\Pi(\omega)\Bigr)
\sim {g^2\over g^2(\omega)} \Biggl\{ 1
+ {1\over\omega^2}\Bigl[
2m^2\bigl(1-{g^2(\omega)\over 2\pi}\bigr)
-4g^2(\omega)\langle[\theta_{00}]\rangle_{T,\mu}\Bigr]
+{\cal O}\bigl({1\over\omega^4}\bigr)\Biggr\} \, ,
\end{equation}
which leads to the following asymptotic form for the continuum
part of the spectral function
\begin{eqnarray}
{1\over N}\rho_{\text{power}}(\omega)
&\sim&{1\over 2}{\bigl(g^2(\omega)/g^2\bigr)^2\over 1+g^4(\omega)/4}
-{g^4(\omega)\over\omega^2}\langle[(\bar{\psi}\psi)^2]\rangle_{T,\mu}
\nonumber \\
&+&{4g^2(\omega)\over\omega^2}\Bigl({g^2(\omega)\over g^2}\Bigr)^2
\langle[\theta_{00}]\rangle_{T,\mu}
\Bigl\{1+{\cal O}\bigl(g^2(Q)\bigr)\Bigr\}\, .
\label{rho_asym}
\end{eqnarray}
Here we only concern with the $g^2(Q)$-dependent terms and postpone
the pure $1/Q^2$ terms (related to $\Delta\rho_{\text{exp}}$) to a
later subsection. We recognize again the first term in Eq.~(\ref{rho_asym}
as the asymptotic perturbative spectral function. Moreover,
the $(T,\mu)$-dependent part of Eq.~(\ref{rho_asym}) has indeed
the general form of Eq.~(\ref{rholn}).

In connection with the comments we made after Eq.~(\ref{rholn}), we point
out that Eq.~(\ref{rho_asym}) has been obtained by expanding the spectral
function in the limit $Q^2\to\infty$ at fixed $T$. Had we made
instead a high-$T$ expansion ($T\to\infty$) at fixed $Q^2$,
we would have obtained, for instance,
$\ln(T^2/\omega^2)$ from the third term of Eq.~(\ref{repi}).

Upon substituting Eq.~(\ref{rho_asym}) in the dispersion integral,
and using the identity
$$
\int \!{d\omega^2\over \omega^2+Q^2}\,
{g^{2\nu}(\omega)\over\omega^2}\sim
{2\pi\over 1-\nu}{g^{2(\nu-1)}(Q)\over Q^2}
\Bigl\{1+{\cal O}\bigl(g^2(Q)\bigr)\Bigr\} \,
$$
where $\nu>1$ and we have kept only the  $g^2(Q)$-dependent part,
we find the following series to the leading order in $g^2(Q)$,
\begin{equation}
\int \!{d\omega^2\over \omega^2+Q^2}\,{\rho(\omega)\over N}
\sim \frac{\langle J_5 J_5\rangle^{(0)}(Q)}{N}
-2g^2(Q){\langle[(\bar{\psi}\psi)^2]\rangle_{T,\mu}\over Q^2}
+ 4\Bigl({g^2(Q)\over g^2}\Bigr)^2
{\langle[\theta_{00}]\rangle_{T,\mu}\over Q^2}+\cdots
\label{gndisint}
\end{equation}
This result is in agreement with
the OPE series of Eq.~(\ref{corrQ}), which has been directly expanded
from the correlator in the Euclidean region.

The fact that no pure $1/Q^2$ term appears in Eq.~(\ref{gndisint}) implies
that the zeroth moment of the subtracted spectral function vanishes:
$\overline{\Delta\rho}=0$. The zeroth moment of the spectral function
without the subtraction is not convergent, because of the contribution
from the perturbative term, even if one can define it through a proper
analytic continuation that yields $\overline{\rho}=0$.

\subsection{Direct solution in terms of the OPE}
We have just verified that in the Gross-Neveu model the exact correlation
function can indeed be expanded in an asymptotic series precisely in
the form of the OPE. Now we shall calculate the Wilson coefficients,
$\beta$-function and appropriate anomalous dimensions for the OPE, and
verify that they match the coefficients and exponents of this asymptotic
series. Furthermore, though it is beyond the scope of the OPE and
RGE, we shall also calculate the condensates.
In this way, we explicitly verify that the asymptotic expansion of the
exact correlation function is identical to the OPE.

\subsubsection{One-loop calculation}
  The one-loop $\beta$-function of the 1+1 dimensional Gross-Neveu
model has already been calculated in the original paper~\cite{gn}:
$\beta=-g^4/2\pi$, i.e. according to our notation $b=1/2\pi$.

  The anomalous dimensions for pseudoscalar current
$J_5\equiv\bar{\psi}i\gamma_5\psi$ and four-quark operator
$O_4=(\bar{\psi}\psi)^2$ can be calculated using the Feynman diagrams
shown in Fig.~\ref{fig4}a and Fig.~\ref{fig4}b, respectively. According
to the notation of Eq.~(\ref{rgef}) we find $\gamma_{J_5}=1/\pi$, and
$\gamma_{O_4}=2/\pi$. The energy-momentum conservation makes
the anomalous dimension for $\theta_{00}$ vanish, i.e.
$\gamma_{\theta_{00}}=0$.

  The relevant Wilson coefficients corresponding to operators $O_4$
and $\theta_{00}$ can be calculated, to leading order in $g^2(Q)$,
using the Feynman diagrams shown in Fig.~\ref{fig5}a and Fig.~\ref{fig5}b,
respectively. The resulting coefficient are
$c_{O_4}=-2g^2(Q)$ ($\nu^{(0)}_{O_4}=1$ according to notation in
Eq.~(\ref{asyope})) and $c_{\theta_{00}}=4$ ($\nu^{(0)}_{\theta_{00}}=0$).

These results lead to the exponents
$\eta_{O_4}=1+(2\gamma_{J_5}-\gamma_{O_4})/2b=1$ and
$\eta_{\theta_{00}}=(2\gamma_{J_5}-\gamma_{\theta_{00}})/2b=2$,
which are precisely what we found in Eq.~(\ref{gndisint}).

  These explicit one-loop calculations exactly match the solution
in Eq.~(\ref{gndisint}).

At last a comment on the one-loop calculation: this result is exact
to the leading order in $1/N$ and only involves
insertion of the quark-bubble chain. Since the quark-bubble behaves
like $\propto\ln Q^2$, this insertion is
equivalent to substituting the coupling constant $g^2$ with the
running coupling constant  $g^2(Q)$. So one can easily identify those
sets of Feynman graphs whose sum leads to the solution of the RGE in
Eq.~(\ref{rge}).

\subsubsection{Condensates}
  The OPE itself does not specify how the relevant condensates are
calculated, since the OPE is only a RGE-improved perturbative
procedure. The calculation of the condensates can be done only in
a non-perturbative context. Here, we carry out this calculation
using the $1/N$ expansion.

  The bare quark condensate $\langle \bar{\psi}\psi \rangle_{T,\mu}$
is given by the dynamical quark tadpole graph
\begin{equation}
\langle \bar{\psi}\psi \rangle_{T,\mu}
=(-iT)\int^\Lambda{dk\over 2\pi} \sum_n\text{Tr}{i\over k.\gamma-m}
=2mT\int^\Lambda{dk\over 2\pi}
\sum_n{1\over (i\omega_n+\mu)^2-(k^2+m^2)}\, .
\end{equation}
Using the standard contour integral technique to carry out the
Matsubara frequency sum, we obtain
\begin{equation}
\langle \bar{\psi}\psi \rangle_{T,\mu}
=m\langle\!\langle E_k^0\rangle\!\rangle
-{m\over\pi}\ln\Bigl({2\Lambda\over m}\Bigr)\, .
\end{equation}
The renormalized quark condensate is obtained by replacing $2\Lambda$
with the subtraction point $\kappa$ in the above equation
\begin{equation}
\langle[\bar{\psi}\psi]\rangle_{T,\mu}
=m\langle\!\langle E_k^0\rangle\!\rangle
-{m\over\pi}\ln\Bigl({\kappa\over m}\Bigr)=-{m\over g^2}\, .
\end{equation}
In the last step we have used the gap equation Eq.~(\ref{gapsh}).
Notice that the renormalization of this condensate is not independent
of the one carried out for the effective potential.
In fact, the $\sigma$-field is just an auxiliary field:
$\sigma=-g^2 \bar{\psi}\psi$. Therefore, the gap
equation gives us not only the expectation value of the $\sigma$-field,
but also of $\bar{\psi}\psi$:
$\langle\sigma\rangle_{T,\mu} \equiv m(T,\mu) =
-g^2 \langle[\bar{\psi}\psi]\rangle_{T,\mu}$. So the renormalizations
of $\bar{\psi}\psi$ and $g^2$ are related.

The four-quark condensate factorizes in the large-$N$ limit, and we find
$\langle[(\bar{\psi}\psi)^2]\rangle_{T,\mu}=
\langle[\bar{\psi}\psi]\rangle_{T,\mu}^2=m^2/g^4$.

  The bare kinetic energy expectation value is given by
\begin{equation}
\langle \bar{\psi}i\gamma_0\partial_0\psi \rangle_{T,\mu}
=(-iT)\int^\Lambda{dk\over 2\pi}
    \sum_n \text{Tr}\gamma_0 k_0{i\over k.\gamma-m}
= 2T\int^\Lambda{dk\over 2\pi}
\sum_n{(i\omega_n+\mu)^2\over (i\omega_n+\mu)^2-(k^2+m^2)}\, .
\end{equation}
After removing the $(T,\mu)$-independent volume and quadratic divergences,
and introduced a subtraction point for the logarithmic divergence
the renormalized kinetic energy becomes
\begin{equation}
\langle[\bar{\psi}i\gamma_0\partial_0\psi]\rangle_{T,\mu}
=\langle\!\langle E_k^2 \rangle\!\rangle
-{m^2\over 2\pi}\ln\Bigl({\kappa\over m}\Bigr)\, .
\end{equation}
After subtracting the trace term from the kinetic energy, we finally
obtain the expectation value of the traceless energy-momentum tensor
\begin{equation}
\langle[\theta_{00}]\rangle_{T,\mu}
\equiv \langle[\bar{\psi}i\gamma_0\partial_0\psi
         -{1\over 2}m\bar{\psi}\psi]\rangle_{T,\mu}
= \langle\!\langle E_k^2\rangle\!\rangle
     -{m^2\over 2}\langle\!\langle E_k^0\rangle\!\rangle\, .
\end{equation}
In the symmetric phase ($m=0$), the expectation value of the
energy-momentum tensor has the very simple form at $\mu=0$
\begin{equation}
\langle[\theta_{00}]\rangle_{T,\mu=0}={\pi\over 6}T^2\, .
\end{equation}

One can verify that inserting these condensates in the OPE, whose
coefficients and exponents we have already calculated, reproduces the
asymptotic expansion of the exact correlator in Eq.~(\ref{corrQ}).

\subsection{Explicit verification of the exact sum rule}
Let us first notice that the sum rule
$\int d\omega^2 \Delta\rho(\omega) =0$
converges very slowly. In fact,
the subtracted spectral function has the leading behavior
\begin{equation}
\Delta\rho(\omega)\equiv\Delta\rho_2(\omega)
             + {\cal O}\bigl({1\over\omega^4}\bigr)\, ,
\end{equation}
where
\begin{equation}
\Delta\rho_2(\omega)
= {4\over\omega^2}{g^4(\omega)\over g^4}
\biggl\{-{\Delta m^2\over g^4(\omega)+4}+{4g^2(\omega)\over\pi}
{\Delta(m^2+4\pi\langle[\theta_{00}]\rangle_{T,\mu})\over
g^4(\omega)+4}\biggr\}\, .
\label{drho2}
\end{equation}
Since $g^2(\omega)=\pi/\ln(\omega/M)$, the leading behavior
in Eq.~(\ref{drho2}) is
$\Delta\rho_2(\omega)\sim [\omega^2(\ln\omega)^2]^{-1}$, and it is clear
that the sum rule only converges logarithmically, i.e.
$\lim_{A^2\to\infty}\int_0^{A^2} d\omega^2 \Delta\rho(\omega)\sim 1/\ln{A^2}$.

Moreover, it is only thanks to the the logarithmic corrections
that the sum rule is finite. Had we disregarded this corrections,
we would have found $\Delta\rho_2(\omega)\sim \omega^{-2}$, and
the zeroth  moment of the subtracted spectral function would not exist.

 Since  the exact spectral function cannot be integrated analytically,
we verify the exact sum rule numerically. We take advantage of the fact
that the integral of $\Delta\rho_2(\omega)$ in Eq.~(\ref{drho2})
can be done analytically
to improve the numerical convergence of the integral. We rewrite the
sum rule as
\begin{equation}
\int_0^{A^2} d\omega^2\,\Delta\rho(\omega)
= -\int_{A^2}^\infty d\omega^2\,\Delta\rho(\omega)
= -\int_{A^2}^\infty d\omega^2\,\Delta\rho_2(\omega)
+{\cal O}\bigl({1\over A^2}\bigr)\, ,
\end{equation}
where the integral of $\Delta\rho_2(\omega)$
can be done analytically and we obtain
\begin{equation}
\int_0^{A^2} \!\! d\omega^2\,\Delta\rho(\omega)
= {\pi\over g^4}\biggl\{2\Delta m^2\biggl({2\over\pi}
\tan^{-1}{2\ln A\over\pi}-1 \biggr)
+ {\Delta(m^2+4\pi\langle[\theta_{00}]\rangle_{T,\mu})\over
(\ln A)^2+\pi^2/4}\biggr\}+{\cal O}\bigl({1\over A^2}\bigr)\, .
\end{equation}
This formula can be easily verified numerically with high accuracy,
since the error is now $1/A^2$, rather than $1/\ln A$. It is reminded
that $\Delta\rho$ in the above equation should be obtained from the
complete SF in Eq.~(\ref{rho_full}), including both
$\Delta\rho_{\text{power}}$ and $\Delta\rho_{\text{exp}}$.

\subsection{Correlator in vector channel and baryon number susceptibility}
There are several reasons to study the correlator in the vector channel
in the Gross-Neveu model.
First, the vector channel gives us an example where the zeroth moment
of the subtracted SF does not exist, but the asymptotic procedure can
still be carried out. Moreover, in connection with the fact that the
vector current is conserved, this spectral function is not affected by
logarithmic corrections, and we can use it to illustrate the use
of the Mellin transform method to obtain the asymptotic expansion.
In addition, very much like in the QCD case, we shall see that the
baryon number susceptibility changes drastically near the chiral
restoration transition, even if we do not expect any real singularity
associated with critical phenomena in the vector channel.
For convenience, we only consider the case of $\mu=0$.

\subsubsection{Vector correlator in Euclidean region}
  The Feynman graphs that contribute to the vector correlator are
depicted in Fig.~\ref{fig6}. Note that if there were no mixing
between vector and pseudoscalar channel, which in fact is only present
in 1+1 dimensions,
the sole contribution would come from the first graph.
The total contribution from all the graphs in Fig.~\ref{fig6} is
\begin{equation}
\langle J_\alpha J^\alpha\rangle_{T,\mu}(iQ) =
 {1\over\pi}\biggl\{1 -{8m^2\over\sqrt{Q^2(Q^2+4m^2)}}
\tanh^{-1}\sqrt{Q^2\over Q^2+4m^2}
+ 2m^2\int_0^\infty{dk\over E_k}\,
{f(E_k;T,0)\over E_k^2+Q^2/4}\biggr\}\, .
\label{vvceucl}
\end{equation}
We define the spectral function in vector channel as
\begin{equation}
\langle J_\alpha J^\alpha\rangle_{T,\mu}(iQ)={1\over\pi}
\biggl\{1-\int_0^\infty d\omega^2
{\tilde{\rho}(\omega)\over \omega^2+Q^2}\biggr\}\, .
\end{equation}
Then we find the following explicit expression for $\tilde{\rho}$
\begin{equation}
\tilde{\rho}(\omega)=
{4m^2\theta(\omega^2-4m^2)\over\sqrt{\omega^2(\omega^2-4m^2)}}
\Bigl(1-{2\over e^{|\omega|/2T}+1}\Bigr)\, .
\end{equation}
It is clear that leading asymptotic behavior of the spectral function is
$\tilde{\rho}(\omega)\sim 1/\omega^2$, which implies that the zeroth
moment of the spectral function does not exist (it is divergent).
In the section~\ref{general} we have shown that, corresponding to
this divergence, we must find negative powers of $g^2(Q)$ (positive power
of $\ln(Q^2)$) in the OPE series.

When there are no explicit logarithms in $\tilde{\rho}(\omega)$, it
is easier to make the asymptotic expansion by using the Mellin
transform method, and in particular the convolution property
of the Mellin transform
\cite{bleistein},
\begin{equation}
\int_0^\infty dt\,f(t)h(t)={1\over 2\pi i}
\int_{c-i\infty}^{c+i\infty}dz\,M[h;z]M[f;1-z]\, ,
\label{convolut}
\end{equation}
where $M[h;z]$ is the Mellin transform of $h(t)$, defined by
\begin{equation}
M[h;z]\equiv\int_0^\infty dt\, t^{z-1}h(t)\, .
\end{equation}
The value of $c$ in Eq.(\ref{convolut}) depends on the asymptotic
behaviors of $h(t)$ at $t\rightarrow\infty$ and $f(t)$ at
$t\rightarrow 0$ respectively; details can be found, for instance,
in Chapter 4 of Ref.~\cite{bleistein}.

  Now let us illustrate how to use this method to expand the
power part of the spectral function (the part containing exponential
dumping factors can be expanded naively)
\begin{equation}
\tilde{\rho}_{\text{power}}(\omega)={4m^2\theta(\omega^2-4m^2)
\over\sqrt{\omega^2(\omega^2-4m^2)}}\, ,
\end{equation}
whose Mellin transform is
\begin{equation}
M[\tilde{\rho}_{\text{power}};z]=
(4m^2)^z\sqrt{\pi}\Gamma(1-z)/\Gamma(3/2-z) \, .
\end{equation}
Using the convolution, Eq.~(\ref{convolut}), and the Mellin transform
of $1/(u^2+Q^2)$
\begin{equation}
M[(u^2+Q^2)^{-1};1-z]=\pi(Q^2)^{-z}/\sin\pi(1-z) \, ,
\end{equation}
we find (in this case $c$ in Eq.(\ref{convolut}) obeys $0<c<1$)
\begin{eqnarray}
\int_0^\infty\,d\omega^2{\tilde{\rho}_{\text{power}}\over\omega^2+Q^2}
&=&-\sum_{n=1}^\infty\,\text{Res}\biggl[\Bigl({4m^2\over Q^2}\Bigr)^z
{\sqrt{\pi}\Gamma(1-z)\over\Gamma(3/2-z)}
{\pi\over\sin\pi(1-z)}\biggr]_{z=n} \nonumber \\
&=& {4m^2\over Q^2}\ln\Bigl({Q^2\over m^2}\Bigr)
-{8m^4\over Q^4}\Bigl[1+\ln\Bigl({Q^2\over m^2}\Bigr)\Bigr]
+{\cal O}\Bigl({1\over Q^6}\Bigr)\, .
\label{asymmell}
\end{eqnarray}
Similarly to the pseudoscalar correlator, it is possible to check
this result in two ways. We can directly expand the power part
(temperature independent part) of Eq.~(\ref{vvceucl}), and we can
derive the OPE ($\gamma_V=0$ and $c_{O_4}=8g^2(Q)$ for the $1/Q^2$
term).

As promised, there indeed appears a positive power of $\ln(Q^2)$
in the OPE series. At the same time, the singularities of
$M[\tilde{\rho}_{\text{power}};z]$ at $z=$ positive
integer hints the fact that all the non-negative integer moments of
$\tilde{\rho}_{\text{power}}$ do not exist. On the other hand, the
asymptotic expansion in Eq.~(\ref{asymmell}) is well defined. Even
though the vector current is conserved, due to the fact that the
anomalous dimension of $O_4$ is not zero, the correct result can not
be obtained by
a naive asymptotic expansion in this case, in accordance with the
remark number (6) at the end of section II.

This same result could have been obtained also by the general method
developed in section~\ref{general}, but it should be clear by now
that the use of the Mellin transform, when possible, is more
straightforward and require less labour.

\subsubsection{Baryon number susceptibility}
The baryon number susceptibility ($\chi_{\text{BN}}$) is defined
as the vector correlator in the limit $Q^2\rightarrow 0$.
{}From Eq.~(\ref{vvceucl}) we find
\begin{equation}
\chi_{\text{BN}}
\equiv \pi\langle J_\alpha J^\alpha\rangle_T(iQ\rightarrow 0)
= -1+2m^2\int_0^\infty{dk\over E_k^3}\,{2\over e^{E_k/T}+1}\, .
\end{equation}
Notice that since we have contracted the
Lorentz indices when defining our vector correlator, $\chi_{\text{BN}}$
is nonpositive definite. It is also possible that our definition
of $\chi_{\text{BN}}$ differs from others by an additive constant,
which is irrelevant to the temperature dependence we are considering.
In the symmetric phase, the limit $m\rightarrow 0$ should be take with
some care: the result is $\chi_{\text{BN}}=1$, when $T>T_c$. The complete
temperature dependence of $\chi_{\text{BN}}$ is displayed in
Fig.~\ref{fig7}b.
As expected, there is no divergence or critical phenomena in this channel.
Nevertheless, we do see a sudden rising of $\chi_{\text{BN}}$ in
correspondence to the relatively rapid drop of the dynamical quark mass,
which we show in Fig.~\ref{fig7}a for comparison,
near the phase transition region.

\section{Application to QCD}
\label{QCD}
  After a lengthy detour to the Gross-Neveu model let us now back
to the main interest of our work, to derive exact sum rules and
to explore the corresponding phenomenological consequences in QCD.
As we will see, QCD shares many of the qualitative features
observed in the Gross-Neveu model, implying the simple fact
that any drastic change induced by a phase transition or rapid
crossover would necessarily be reflected through the spectral
function in proper channels at low frequencies, independent of
the details of the model.

\subsection{Sum Rules}
  Let us consider four correlators of mesonic currents in QCD.
The correlation between scalar currents and the one between
pseudoscalar currents involve non-conserved operators,
$J_S=\bar{\psi}\psi$ and $J_P=\bar{\psi}\gamma_5\psi$, whose anomalous
dimensions are $\gamma_{J_S}=\gamma_{J_P}=1/4\pi^2$. One the other end,
correlators between vector currents and axial-vector currents involve
conserved quantities, $J_V=\bar{\psi}\gamma_\mu\psi$ and
$J_A=\bar{\psi}\gamma_\mu\gamma_5\psi$, whose anomalous dimensions
vanish ($\gamma_V=\gamma_A=0$).

In QCD there exist several dimension-four operators, but all of
them have non-positive anomalous dimensions, therefore the two
correlators between nonconserved currents have
$\eta_n\geq (2\gamma_J-\gamma_n)/2b>0$ and Eq.~(\ref{nonc}) applies,
{\em i.e.} the zeroth moments of their SF's are independent of $T$
and $\mu$.

On the other hand, the two conserved currents have, in correspondence
with conserved operators
$\eta_n=0$ and a generalization of Eq.~(\ref{cons}) applies.
In this case in fact there are
three dimension-four operators with zero anomalous dimension.
Two of these operators are Lorentz scalars:
$m\bar{\psi}\psi$ and
$\alpha_s G^2\equiv\alpha_s \text{tr}(G_{\mu\nu}G^{\mu\nu})$,
while the third is the energy-momentum tensor:
\begin{equation}
\theta_{\mu\nu}=
 \frac{1}{4} \bar{\psi}(\gamma_{\mu} i\tensor{D}_{\nu} +
 \gamma_{\nu} i\tensor{D}_{\mu}-m g_{\mu\nu}) \psi
+2\text{tr}(G_{\mu\alpha}G^{\phantom{\nu}\alpha}_{\nu})
-\frac{1}{2} g_{\mu\nu}G^2 \, .
\end{equation}
The sum rules for the vector and axial-vector currents
(with the Lorentz indices contracted) are:
\begin{equation}
\int_0^\infty\!\! du^2\, \Delta\rho(u)=
a\, \Delta\langle[m\bar{\psi}\psi]\rangle
+{\Delta\langle[\alpha_s G^2]\rangle\over 2\pi}
+8\Delta\langle [\theta_{00}]\rangle\, ,
\label{va}
\end{equation}
with $a=6$ and $a=-10$ for vector and axial-vector cases, respectively.

  Obviously, the general remarks at the end of section II apply to
the sum rules derived above. However, there is a new feature of QCD
which may not be shared by all asymptotically free theories. It is
known that QCD OPE series has explicit instanton induced corrections
that cannot be related to condensates of some local operators. But
these exact sum rules are not explicitly affected by this problem,
although the value of the condensates certainly have instanton
contributions. The reason is that the instanton singularities in the
Borel-plane are located on the positive axis starting at $8\pi^2$,
which in turn implies that the explicit contribution of instantons
to correlation functions are of higher order than $1/Q^2$~\cite{thooft}
in the OPE series.

\subsection{Phenomenological Consequences}
Finally, let us discuss some of the phenomenological consequences
of these exact sum rules. The power of these exact sum rules rests
on their generality, in the sense that there is no assumptions on
the functional form of the SF's and they can be applied anywhere
in the $(T,\mu)$-plane.

First let us consider the pseudoscalar channel. The sum rule
$\overline{\Delta\rho}=0$ implies that, in the broken-chiral-symmetry
phase, the change of the pion pole induced by $T$ or $\mu$
is exactly compensated by a corresponding change of the continuum
part of the SF.

Next let us consider the scalar correlation function at zero
frequency, i.e. the chiral susceptibility (a measure of the
fluctuation of the chiral order parameter),
\begin{equation}
\chi(T,\mu)\equiv
\int\! d^4x\, \theta(x_0)\langle [J_S(x),J_S(0)]\rangle_{T,\mu}
=\int_0^\infty\!\! du^2\, {\rho(u;T,\mu)\over u^2}\, ,
\label{chisus}
\end{equation}
which diverges when $(T,\mu)$ approaches the phase boundary,
provided the chiral restoration is a continuous transition.
The divergence of the chiral susceptibility near the phase transition
can only be produced in Eq.~(\ref{chisus}) by singularities very close
to the origin; singularities not near the origin are not compatible
with the sum rule $\overline{\Delta\rho}=0$.
Thus, when approaching the phase boundary in the $(T,\mu)$ plane,
the threshold of the spectral function vanishes (since there is no
massless pole in the chirally symmetric phase), and a strong peak
develops right above threshold. In the chirally symmetric phase the
pseudoscalar and scalar channels are degenerate, therefore we expect
the same behavior for the pseudoscalar SF. This strong
peak in the pseudoscalar and scalar SF's, which is intimately
connected with critical phenomena of a diverging susceptibility
and correlation length near the phase transition,
can be interpreted as some kind of quasi-particle, thus
confirming the qualitative picture, originally proposed in the
context of the Nambu-Jona-Lasinio model~\cite{kunihiro},
of the appearance of soft modes near the chiral phase transition.

We wish to remark that the sum rule has been used only to exclude the
logical possibility of the SF developing singularities at finite energies.
But the sum rule is not a necessary condition for the appearance of a peak.
The presence of a second order phase transition (infinite correlation
length) is the real physical cause of the peak. Nevertheless, one should
not assume that if the phase transition is not second order the
zero-$T$ or perturbative spectral function is a good approximation to the
real spectral function. Even if the chiral restoration turns out to be
a cross-over or weak first order transition (finite but large correlation),
as the lattice data seem to indicate~\cite{phase}, we still expect the
same qualitative features, though less pronounced: a peak develops but
it does not actually diverge at the transition.
Our calculation in the Gross-Neveu model is a clear illustration of
this situation. While Fig.~(\ref{fig2}) shows the divergent peak
of the spectral function near the second order boundary, we see in
Fig.~(\ref{fig3}) that the spectral function is also strongly peaked
near the first order transition as long as we are not too far away from
the ``tricritical point''.

We believe that similar results hold in the vector and axial-vector
channels, even if our conclusions cannot be as strong as in the
previous case, because of two main differences. The first is that
now $\overline{\Delta\rho}\neq 0$. This problem is not very serious,
since we only need that $\overline{\Delta\rho}$ is not singular
crossing the phase boundary. And this can still be argued by means
of the sum rule in Eq.~(\ref{va}) and the fact that, for a
continuous phase transition, the changes of the thermal energy
$\Delta\langle [\theta_{00}]\rangle$, and of both condensates
$\Delta\langle [m\bar{\psi}\psi]\rangle$ and
$\Delta\langle[\alpha_s G^2]\rangle$ should behave smoothly across
the critical line. The most serious difference is that, in these
channels, we cannot argue on physical grounds that the corresponding
susceptibility diverges. Nonetheless,  there exist lattice
simulations~\cite{gottlieb} showing that the so-called quark number
susceptibility rapidly increases in the transition. Thus it is plausible
that also in the vector channel the SF has a vanishing threshold, and
accumulates strength just above it. A similar argument in the context
of the Nambu-Jona-Lasinio model can be found in Ref.~\cite{kuni}.

\section{Summary and outlook}
\label{conclusion}
  We used the operator product expansion and the renormalization
group equation to derive exact sum rules at finite $T$ and $\mu$
valid for asymptotically free theories. Our derivation explicitly
show that logarithmic corrections cannot be neglected.
We found that, depending on the theory and on the current under
study, the zeroth moment of a spectral function is either independent
of $T$ and $\mu$, or its change is related to the corresponding changes
of the condensates of operators of lowest dimension.

In particular the zeroth moment of the scalar and pseudoscalar mesonic
currents in QCD are independent of $T$ and $\mu$. As a consequence any
change in the strength of the pion pole must be exactly compensated
by a change in the continuum contribution. We also infer that spectral
functions in the scalar and pseudoscalar channel should dramatically
change near phase transition. We also find that the zeroth moment of
the vector and axial-vector mesonic currents in QCD changes with $T$
and $\mu$, and these changes are related to the corresponding changes
of the condensates.

Due to their generality, these exact sum rules strongly constrain
the qualitative shape of SF's, in particular, near phase transitions.
Since it appears that a strong deformation of the spectral function
from its counter part at $(T,\mu)=0$ and perturbative cases is
the most likely scenario near phase transitions, the information
carried by these exact sum rules is particularly welcome. We
urge whoever parameterizes a spectral function, e.g.
in QCD sum rule type of calculations or to interpret lattice
simulation data, to incorporate these exact constraints.

We have also illustrated in great detail the derivation
of the sum rules, and confirmed their validity,
in the Gross-Neveu model, where we can also calculate
the exact spectral function. Even if this model calculation is meant
mostly as an illustration of many delicate issues of the derivation
(such as convergence, asymptotic expansion, Mellin transform, high
momentum expansion vs. high temperature expansion, etc.), it is
nevertheless comforting to find that all our general expectations about
the spectral function are explicitly verified in the model.

In the future, we plan to generalize these results
to baryonic currents and to analyze the phenomenological
consequences of these sum rules in greater detail.

This work is supported in part by funds
provided by the U.S.~Department of Energy (D.O.E.) under contracts
\#DE-AC02-76ER03069 and \#DE-FG06-88ER40427.

\begin{figure}
%\vspace{1cm}
\caption{Phase diagram of the Gross-Neveu model in 1+1
dimension. The continuum line represent the second order
phase boundary in the $(T/T_c,\mu/M)$-plane, while the dashed line
is the first order phase boundary. The ``tricritical point'', where
first and second order transitions merge, is shown as a heavy dot.
The dotted line separates the values of $(T,\mu)$ for which the
symmetry is always broken from those values of $(T,\mu)$ for which
a ``metastable symmetric phase'' is possible. Dots show those values
of  $(T,\mu)$ at which we plot the spectral function in
Fig.~\protect\ref{fig2} and Fig.~\protect\ref{fig3}.
Their coordinates are: $a=(0.352,0.4)$, $b=(0.848,0.4)$,
$c=(0.875,0.4)$, $d=(1.763,0.4)$, $e=(1.175,0.4)$, $f=(1.763,0.4)$,
$g=(0.353,0.63)$, $h=(0.353,0.65)$ and $k=(0.353,0.85)$.}
\label{fig1}
\end{figure}

\begin{figure}
%\vspace{1cm}
\caption{Spectral function of the 1+1 dimensional Gross-Neveu model
in pseudoscalar channel. We show several values of $(T,\mu)$ that go
from the broken phase to the symmetric phase passing through the second
order phase transition. The labels a, b, c, d, e and f correspond to
the labels on the points in the $(T/T_c,\mu/M)$-plane shown in
Fig.~\protect\ref{fig1}.
The upper panel shows spectral functions in the broken phase.
The pion poles are indicated by arrows, and are slightly displaced
from their true position at $\omega=0$ for vision convenience.
The length of arrows represents the magnitude of the corresponding
residues. The lower panel shows spectral functions in the symmetric
phase, where there is no pion pole.}
\label{fig2}
\end{figure}

\begin{figure}
%\vspace{1cm}
\caption{The notations are the same as in Fig.~\protect\ref{fig2},
except that now $(T,\mu)$ goes through the first order phase boundary.
The coordinates of the labels a, g, h and k are given in
Fig.~\protect\ref{fig1}.}
\label{fig3}
\end{figure}

\begin{figure}
%\vspace{1cm}
\caption{Feynman diagrams for calculating the anomalous dimensions of
(a) the composite operator $J_5$, i.e. $\gamma_{J_5}$,
(b) and of the composite operator $O_4$, i.e. $\gamma_{O_4}$.}
\label{fig4}
\end{figure}

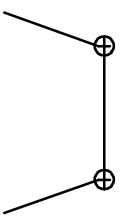
\begin{figure}
%\vspace{1cm}
\caption{Feynman diagrams for calculating Wilson coefficients:
(a) $c_{O_4}$ and (b) $c_{\theta_{00}}$.}
\label{fig5}
\end{figure}

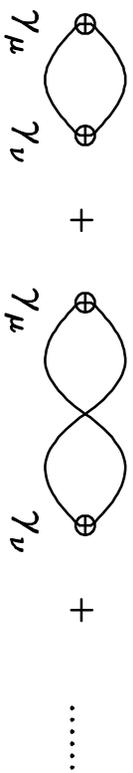
\begin{figure}
\caption{Feynman diagrams contributing to the vector correlator.}
\label{fig6}
\end{figure}

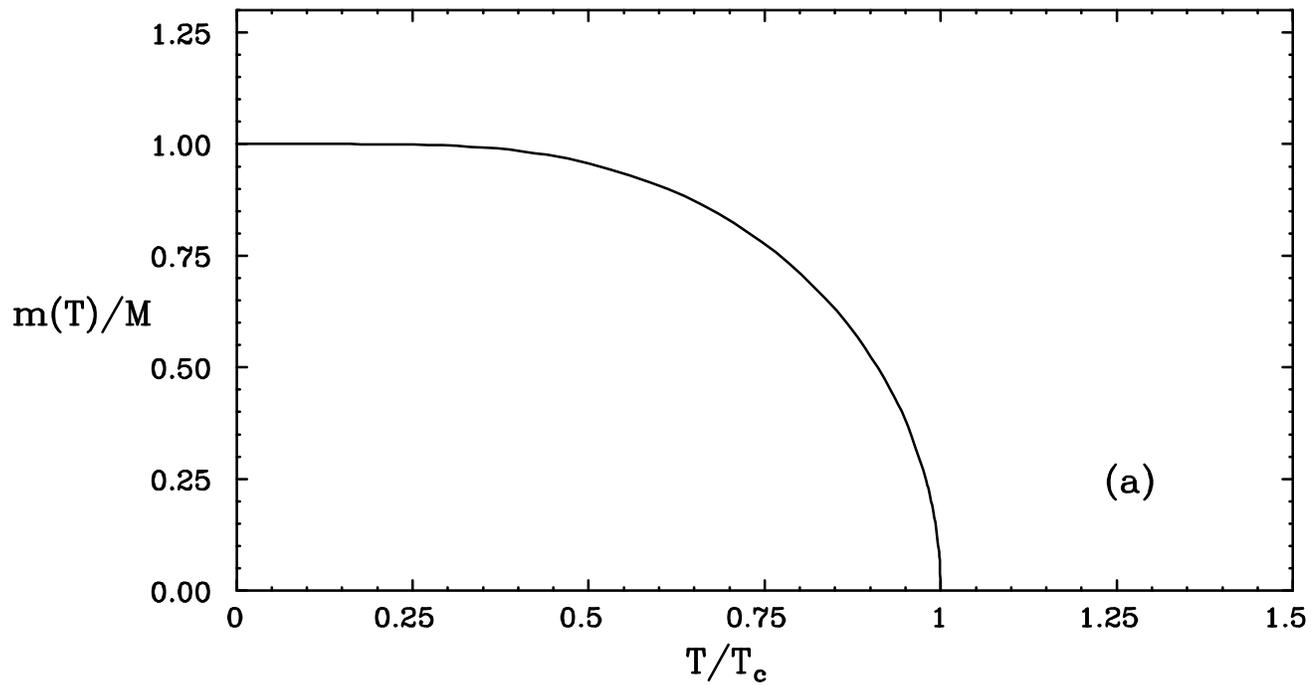
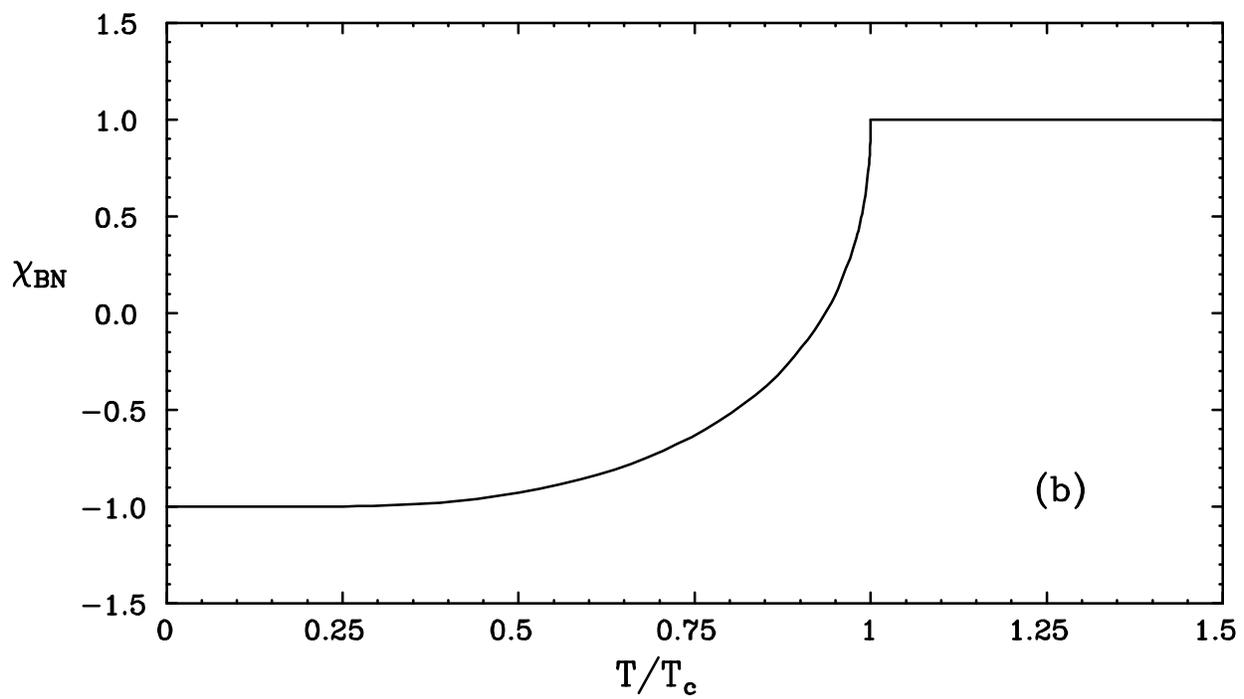
\begin{figure}
\caption{(a) the dynamical quark mass and (b) the baryon number
susceptibility as functions of temperature.}
\label{fig7}
\end{figure}

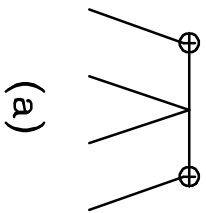
\end{document}